\documentclass[aip,apl,numerical,reprint]{revtex4-1}

\usepackage[dvips]{graphicx}
\usepackage{amssymb}

\newcommand{\eps}{\varepsilon}

\begin{document}

\title{Localized Joule heating produced by ion current focusing through micron-size holes}
\author{V. Viasnoff}
\email{virgile.viasnoff@espci.fr}
\author{U. Bockelmann}
\affiliation{Nanobiophysics Lab, ESPCI, CNRS, 10 rue Vauquelin, 75005 Paris, France.}
\author{A. Meller}
\affiliation{Department of Physics and Biomedical Engineering, Boston University, Boston,
MA 02215, USA.}
\author{H. Isambert}
\affiliation{Institut Curie, CNRS, 11 rue P.M. Curie, 75005 Paris, France.}
\author{L. Laufer}
\author{Y. Tsori}
\affiliation{Department of Chemical Engineering, Ben-Gurion University of the Negev,
84105 Beer-Sheva, Israel.}

\date{\today}

\begin{abstract}
We provide an experimental demonstration that the focusing of ionic currents in a micron
size hole connecting two chambers can produce local temperature increases of up to
$100^\circ$ C with gradients as large as $1^\circ$ K$\mu m^{-1}$. We find a good
agreement between the measured temperature profiles and a finite elements-based numerical
calculation. We show how the thermal gradients can be used to measure the full melting
profile of DNA duplexes within a region of $40$ $\mu$m. The possibility to produce even
larger gradients using sub-micron pores is discussed.
\end{abstract}
\maketitle

The creation of local heat sources and large thermal gradients in confined aqueous
environments is a challenging problem due to the rapid diffusion of heat in water.
Several solutions were proposed, such as heating micro/nanoparticles \cite{HamadNature02,
LevyJPCM08} by magnetic induction or by using a focused laser beam
\cite{BraunePRL02,BaaskeAPL07}. Local thermal gradients in microchannels were used to
sort and concentrate molecules. Several approaches were developed such as thermophoresis
\cite{DuhrPNAS06}, Temperature Gradient Focusing \cite{RossAnalChem02}, Field Gradient
Focusing \cite{KoeglerBioPro96}, and isoelectric Focusing
\cite{PawliszynJMS93,KatesEl06}. These techniques use high DC voltages (tens to hundreds
of Volts) and thermal gradients in the range of  $0.01^\circ$ K$\mu$m$^{-1}$. The typical
volumes are micro to nanoliters. In this letter, we show that a few tens of picoliters
can be strongly heated by focusing an ionic current through a micron size hole in a
saline solution. The resulting gradients are of the order of $1^\circ$ K$\mu$m$^{-1}$.

\begin{center}
\begin{figure}[ht!]
\includegraphics[scale=0.57,clip]{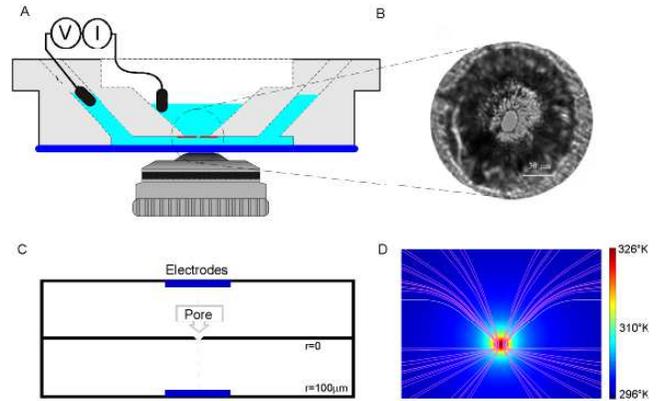}
\caption{A: Schematic representation of the experimental setup. The two chambers are
separated by a $50$ $\mu$m thick Teflon septum punctured by a $7.5$ $\mu$m radius conical
hole B: Close up on the hole region. C: Model set up used for our finite element
modeling. D: close up on the pore region. The ion current lines are represented and the
map of the temperature profile is color coded.}
\label{fig1} 
\end{figure}
\end{center}

We use a custom made cell composed of two chambers separated by a $50$ $\mu$m thick
Teflon septum (see Figure \ref{fig1}). A conical hole ($40^\circ$ half angle, minimum
diameter $r_0=7.5$ $\mu$m) is punctured in the center of the septum. The AC voltage
($\simeq 100V_{pp}$, 10 kHz) is applied across the chambers using platinum electrodes. We
use tris buffer, $1$M KCl, pH 7.4, of electrical conductance $\sigma=107$ mS/cm. The DC
value of the electrical resistance $R_h$ across the hole is $2300~\Omega$. The septum
capacitance is estimated to $C_s=0.4$ pF. It results that the current across the chambers
is mostly resistive at $10$ kHz. The Joule heating power $P_j$ dissipated in the hole is
proportional to the mean root square current $i_{\rm rms}$. We measure the local
temperature profiles along the vertical hole axis in the lower chamber. The temperature
is derived from the confocal detection ($\lambda=532$ nm) of the calibrated fluorescence
of TetraMethylRhodamine (TMR) grafted at the 5'end of DNA oligomers.

\begin{center}
\begin{figure}[ht!]
\includegraphics[scale=0.55,clip]{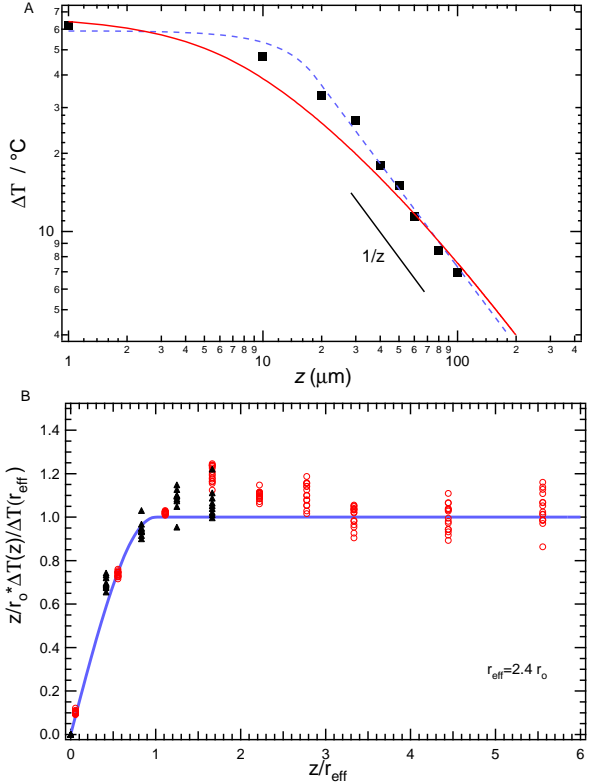}
\caption{A: Profile of the temperature increase $\Delta T(z)$ along the
vertical axis below the hole center. The pore's radius is $7.5$ $\mu$m and the current is
$i_{\rm rms}=1.81$ mA. The average thermal gradient over the first 40 $\mu$m is $1^\circ$
K$\mu$m$^{-1}$. At distances larger than the hole radius the temperature profiles $\Delta
T(z)$ scale as $1/z$. The dotted line is the fit obtained when modeling the heating
source by a spherical source of uniform heating power $P_j=5.2\times 10^{-3}$ W and
radius $r_{\rm eff}=17.8$ $\mu$m. The solid line is the finite element calculation. B:
Rescaling of all temperature profiles obtained with a hole of $7.5\mu$m (open circles)
and $20\mu$m (full triangles) for various currents $(0.5 mA<i_{\rm rms}<1.85 mA)$. We
used $r_{\rm eff}=2.4r_0$. The straight line corresponds to the scaling of the analytical
model. Inset: Values of $\Delta T(r_{eff})$ as a function of $I_{rms}$ for the hole of
$7.5\mu$m (open circles) and $20\mu$m (full triangles).}
\label{fig2}
\end{figure}
\end{center}

Large enough currents, $i_{\rm rms}> 1.9$ mA, result buffer vaporization. Smaller values
of $i_{\rm rms}$ lead to stationary thermal profiles within a few seconds. Figure
\ref{fig2} A shows the temperature increase $\Delta T=T-T_\infty$, where $T_\infty=298
^\circ$K is the room temperature, along the vertical axis for $i_{\rm rms}=1.81$ mA. Over
the first $40$ $\mu$m , the average temperature gradient is $1^\circ$ K$\mu$m$^{-1}$. As
a first approximation we model the heating power $P_{\rm eff}$ as being distributed
uniformly in an effective sphere of radius $r_{\rm eff}$. The temperature then reads:
\begin{eqnarray}
\Delta T(r)&=&\frac{P_{\rm eff}}{8\pi\kappa r_{\rm
eff}}\left(3-\frac{r^2}{r_{\rm eff}^2}\right)
~~~~~r<r_{\rm eff}\nonumber\\
\Delta T(r)&=&\frac{P_{\rm eff}}{4\pi\kappa r} ~~~~~~~~~~~~~~~~~~~~~~~r\geq r_{\rm eff}
\label{ana_formula}
\end{eqnarray}
where $r$ is the distance from the hole center and $\kappa=0.6$ Wm$^{-1}$K$^{-1}$ is the
thermal conductivity of water. Figure \ref{fig2}A shows the measured temperature profile
for hole radius $r_0=7.5$ $\mu$m and power $P_j=6.5\times10^{-3}$ W. The best fit
parameters for our model are  $r_{\rm eff}\simeq 17.8$ $\mu$m and a $P_{\rm
eff}=5.2\times 10^{-3}$ W. This model predicts that when $r>r_{\rm eff}$ the value of
$(r/r_0)\Delta T(z)/\Delta T(r_0)$ is a constant independent of $r_0$ and $P_j$. Figure
\ref{fig2}B shows that the curves for several values of $r_0$ and $P_j$  can be scaled
provided that we take $r_{\rm eff}=2.4r_0$.

In order to explore the influence of the pore geometry on thermal profiles we use a
finite elements approach with various hole radiuses $ r_0$ and pore lengths $L$ . The
numerical calculation accounts for mass transport (\ref{MT}), electrostatic potential
(\ref{CC}) and heat dissipation (\ref{EC}) in the following way:
\begin{eqnarray}
\partial _t C^\pm-\nabla(D^\pm\nabla C^\pm\pm ez\mu^\pm C^\pm\nabla \phi) &=& 0
~,\label{MT}\\
\partial _t (\nabla (\eps \nabla\phi)+\nabla(\sigma \nabla\phi) &=& 0 ~,\label{CC}\\
\rho C_p \partial _t T-\nabla(\kappa \nabla T)-\sigma (\nabla\phi)^2 &=& 0 ~,\label{EC}
\end{eqnarray}
$C^\pm$, $D^\pm$ and $\mu^\pm$ are the number densities, the diffusion coefficients and
the electrophoretic mobilities of the positive and negative ions, respectively. $\phi$ is
the electric potential, $\eps$ is the local dielectric constant, $\sigma$ is the
electrical conductivity, $e$ is the electron's charge, and $C_p$ is the heat capacity.
The steady-state solutions were obtained with an AC voltage at the electrodes. See
Supplementary Material for details.
\begin{center}
\begin{figure}[ht!]
\includegraphics[scale=0.55,clip]{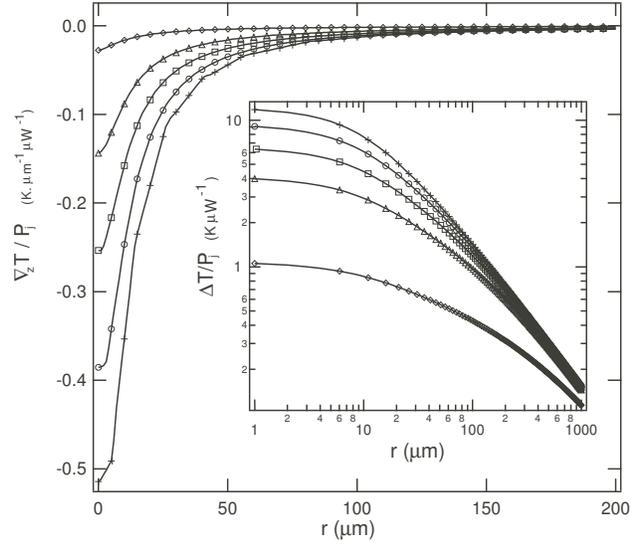}
\caption{Steady state temperature gradients and profiles (inset) below the narrowest
constriction of the hole given by finite elements calculation. $\nabla_z T(z)$ and
$\Delta T(z)$ are scaled by the average heating power $P_j$ for various ratios of hole
size $r_0=10$ $\mu$m with the membrane thickness $L$. The values of $r_0/L$ are $0.01$
($\diamond$), $0.05$ ($\vartriangle$), $0.1$ ($\square$), $0.2$ ($\circ$), $0.5$ (+).
$\Delta T(z)$ agrees with Eq. \ref{ana_formula} and decays like $z^{-1}$ at $z>r_0$ for
$r_0/L$ larger than $0.1$. For lower ratios the thermal gradients are less steep.}
\label{fig3} 
\end{figure}
\end{center}

Figure \ref{fig3} shows the calculated temperature  $\Delta T(z)$ and gradient profiles
$\nabla_z T(z)$ along the z axis at a fixed heating power $P_j$ for several values of the
aspect ratio $r_0/L$. When $r_0/L>0.1$ the temperature is well described by $\Delta
T(z)\sim z^{-1}$  for $z$ values larger than $2r_0$. In the limit of small aspect ratio
the heating is mainly localized in the hole and smoothly spreads in the lower chamber
producing smaller temperature gradients. We conclude that $r_0/L$ should ideally be
between $0.1$ and $1$ to obtain the sharpest gradients and largest temperature increases.

\begin{center}
\begin{figure}[ht!]
\includegraphics[scale=0.62,bb=-0 0 380 280,clip]{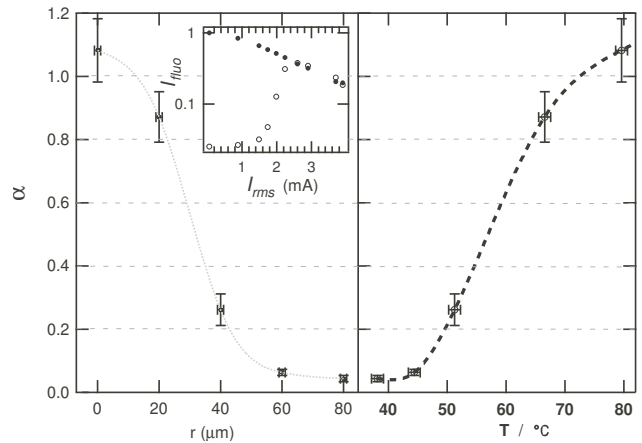}
\caption{The dissociation ratio $\alpha \equiv I_{\rm fluo}({\rm DNA+quencher})/I_{\rm
fluo}({\rm DNA~only})$  as a function of distance and temperature for $i_{\rm rms}=2.9$
mA. A decrease of $90$\% is obtained over $45$ $\mu$m or $35^\circ$ C. The hole diameter
is $40$ $\mu$m. Inset: fluorescence intensity as a function of the ionic current for a
single DNA strand labeled labeled with TMR only (full symbol) and a DNA duplex with TMR
quenched with IowaBlack (open symbol). The lines are guides to the eyes.
}
\label{fig4} 
\end{figure}
\end{center}

The experimental thermal gradients generated by a hole of diameter $r_0=20$ $\mu$m are
used to determine the melting profile a DNA duplex: strand 1: TMR
$5$'-TCAGACCG(TC)$_{15}$-$3$', strand 2: $5$'-CGGTCTGA-$3$' IowaBlack. The DNA was gel
purified to obtain a $95$\% hybridization efficiency. The fluorescence intensity of the
TMR is quenched  $20$-fold on average upon hybridization with IowaBlack. With a proper
baseline calibration the fluorescence intensity measured at the laser spot can be used to
determine the fraction of hybridized duplexes. We measure the fluorescence profiles at
various values of $i_{\rm rms}$ for strand $1$ only, and for the hybridized duplex with a
$1$:$1$ ratio of both strands (see Figure \ref{fig4}). Assuming (i) a local thermal
equilibrium, (ii) a two state model where the 8 mers are either fully hybridized or
completely open, and  (iii) an efficient quenching for all temperatures, we can extract
the dissociation coefficient $\alpha$ of the hairpin as a function of temperature and
distance from the pore. Figure \ref{fig4} shows the full melting profile obtained over
a distance of 45 $\mu$m for a hole of radius $r_0=20$ $\mu$m. Following the typical
melting curve analysis for bimolecular equilibrium \cite{MergnyON03}, we extract the
thermodynamical melting parameters of the DNA structure. We find $\Delta H=-66\pm 6$ kCal
mole$^{-1}$ and $\Delta S=182\pm 25$ Cal mole$^{-1}$ K$^{-1}$, in good agreement with the
thermodynamical parameters calculated with MFold \cite{Mfold} under similar salt
conditions ($\Delta H=-62$ kCal mole$^{-1}$ and $\Delta S=169$ Cal mole$^{-1}$ K$^{-1}$).
Spatial temperature gradients have the advantage over traditional melting curves
techniques that all temperatures can be probed simultaneously. This method works if the
diffusion and/or drift of the DNA molecule across the thermal profile is slow enough to
allow thermal equilibration. As derived in the Supplementary Material, electrophoretic
and electroosmotic drifts are negligible in our experiments. The Brownian diffusion
coefficient \cite{TinlandMacro97} for our DNA molecules is of order $D_{\rm DNA} \simeq
10^{-7}$ cm$^2$s$^{-1}$. The distance over which the temperature changes by $1^\circ C$
is d=$1$ $\mu$m. The diffusion time across this distance is $\tau=d^2/D_{\rm DNA}=0.01$
s. Since small molecular beacons are reported to open over a characteristic time of
$10^{-4}$ s \cite{BonnetPNAS98}, the approximation of local equilibrium is satisfied.

We briefly discuss the influence of various parameters on the applicability of our
device. For frequencies $\omega< (R_hC_s)^{-1}=ar_0^2/(\sigma C_s)$, where $a$ is a
geometry-dependent factor, the ionic current is mostly resistive. It is reasonable to
assume that electroosmotic flow are negligible due to the use of AC fields.
Electroosmosis can be significantly enhanced in smaller ($\simeq 100$ nm) or charged
pores when current rectification occurs \cite{Siwy06}. In the lower chamber natural
convection is minimized since the hot spot (the hole) is located above the cold region.
In addition, there is negligible fluid transfer between the upper and lower chambers. The
small extent of the lower chamber also increases the instability threshold for natural
convection. This situation contrasts with focused laser heating for which convective
rolls are observed along the optical axis \cite{BraunePRL02}. Convection can be
externally applied to drive the molecule through the thermal profile. Our conclusions
would still hold in the limit of small Peclet numbers i.e if the convective velocity is
small: $v<\kappa/\rho C_pr_0=10^{-5}/r_0$ m s$^{-1}$.  In this limit thermal quenching
rate of $1^\circ$ K/$\mu$s can still be achieved. See Supplementary Material for a
detailed discussion. We believe that this approach facilitates the creation of large
thermal gradients in sub-micron regions with potential applications for fast denaturation
and thermal quenching \cite{ViasnoffNanoLett06} to study local chemical reactions.

%

\end{document}


{\large \textbf{Supplementary Material}}

\textbf{The melting profile}

To measure the melting curves of DNA by spatially scanning the thermal profile, we
assumed that the DNA was at thermal equilibrium at each point of space. This assumption
is true if thermal equilibration is fast compared to the drift and diffusion velocity.
The drift can originate from the electrophoretic mobility of the DNA or its convection in
a
electroosmotic flow (if present). Because we use AC voltages the average value of the
drift is zero. Notice that most of the melting profile is measured
outside the hole region where the electric field is small; only the small $z$-values
correspond to the vicinity of the hole where the field is appreciable. 
The excursion of the molecule in the temperature profile during
one period can be evaluated as follows. The
electrophoretic mobility of DNA is largely
independent on the DNA size. Its value \cite{Stellwagen02} is $3\times10^{-4}$ cm
V$^{-1}$ s$^{-1}$.  The maximum electric field can be estimated by dividing the maximum
voltage by the characteristic size, that is $E=25/50\times 10^{-6}=5\times 10^5
V/m$. The
electrophoretic velocity of the DNA in the hole is then equal to
$1.5\times 10^{-2}$ ms$^{-1}$.  Hence the drift distance during half a
period of the field is of order $1$ $\mu$m. At larger distances from the pore the
electric field decreases  as $1/r^2$ and the drift distance is correspondingly decreased.

We could not detect any electroosmotic flow. If any such drift existed it would have been
localized at the hole where all flow lines converge.
We used a Teflon septum in a 1M KCl solution. This means that the zeta
potential and therefore the electroosmotic flow are largely reduced. The electroosmotic
coefficient \cite{Camilleri98} for PTFE
(Teflon) capillary tubing is below $10^{-4}$ V$^{-1}$ s$^{-1}$ rendering this drift 
negligible.

Lastly we point out that the temperature profile is also estimated using
DNA molecules with similar electrophoretic mobility and undergoing the same
electroosmotic flow. Hence, the temperature that we measure at a fixed point is also the
mean temperature over the excursion of the DNA in the temperature profile.

\textbf{Applicability of the device}

This device can be a useful tool to produce large temperature gradients. Several
parameters may affect the applicability of the concept:

1 - \emph{AC frequency}. Heating occurs at all frequencies and increases with decreasing
frequency. At 
small frequencies, however, the drift of the DNA molecules in one period is diminished.
We tested frequencies between 1 to 30 kHz without any significant change
in the results. At large enough frequencies the current across the
membrane can be dominated by the capacitive charging of the membrane. The crossover
frequency is $\omega = (R_hC_s)^{-1}=ar_0^2/(\sigma C_s)^{-1}$, where
$R_h\simeq 2300~\Omega$ and $r_0$ are the pore's resistance and radius, $C_s\simeq0.4$
pF is the septum capacitance, $\sigma$ is the
electrical conductivity and $a$ is a geometry-dependent factor. For the $50$ $\mu$m thick
membrane we used the crossover frequency is $1$ GHz. For thin membranes capacitive
charging should be considered and in this case the ion flux 
occurs everywhere in the solution and the heating is consequently less localized and 
more homogeneous. Note however that the product $R_hC_s$ does not
dependent on the membrane thickness for a planar capacitor and a cylindrical hole.

2- \emph{Buffer composition}. In the Teflon membranes we used the values of pH and
salt
concentration do not dramatically affect the zeta potential at the pore surface. 
Other polymers or solid state membrane can be affected by this parameters. 

As stated above, heating occurs within a few tens of microns of the pore and 
the Peclet number is small. In this limit, the analysis and heating
profile should be unchanged even in the presence of convection. No changes were detected 
in our experiments with solutions of concentrations in the range from 0.2 M to 1M.

\textbf{Numerical calculations}

The governing equation for the salt concentration $C^\pm$, electric potential $\phi$, and
temperature $T$
\begin{eqnarray}
\partial _t C^\pm-\nabla(D^\pm\nabla C^\pm\pm ez\mu^\pm C^\pm\nabla \phi) &=& 0
~,\label{MT}\\
\partial _t (\nabla (\eps \nabla\phi)+\nabla(\sigma \nabla\phi) &=& 0 ~,\label{CC}\\
\rho C_p \partial _t T-\nabla(\kappa \nabla T)-\sigma (\nabla\phi)^2 &=& 0 ~,\label{EC}
\end{eqnarray}
were solved using a finite element method with a high resolution mesh near the pore.
$C^\pm$, $D^\pm$ and $\mu^\pm$ are the concentration, diffusion coefficient and mobility
of positive (``+'', K) and negative ions (``-'', Cl) in water, respectively. 
The diffusion of K$^+$ and Cl$^-$ ions were taken to be $D^+=1.9\times 10^{-9}$ m$^2$/s
and 
$D^-=2\times 10^{-9}$ m$^2$/s and the mobilities depended on the diffusion
via Einstein's relation. $e$ is the electron's
charge and the valency $z$ was chosen
to be $z=1$. The first equation was calculated in the aqueous region only as ions cannot
penetrate the hard walls. 
The second equation was solved in the aqueous regions with electrical
conductivity given by $\sigma=e^2(\mu^+C^++\mu^-C^-)$ and dielectric constant
$\eps=80\eps_0$, where $\eps_0$ is the vaccum permittivity, and with $\sigma=0$
and $\eps=2.1\eps_0$ in the solid parts.
The usual boundary conditions at the interfaces between dielectric or
conductive materials were used.

Due to the total electro-neutrality and the AC fields we used, convection
can be neglected and the Navier-Stokes equation need not be solved. Subsequently
the heat equation (third equation) has only Joule heating terms and no viscous
dissipation. $C_p$, $\kappa$ and $\rho$ are the specific heat, heat conductivity, and mass
density of the aqueous region and Teflon cell. We used $C_p=4180$ J/Kg
K$^\circ$, $\kappa=0.58$ W/mK$^\circ$, and $\rho=1000$ Kg/m$^3$ in water and 
$C_p=1000$ J/Kg K$^\circ$, $\kappa=0.27$ W/mK$^\circ$, and $\rho=2140$ Kg/m$^3$
in Teflon.
At the water-Teflon interface we conserved
the heat flux, and at the edges of the cell we used Newtonian heat loss to the environment
(proportional to $T-T_\infty$) with temperature at infinity $T_\infty$ equal to the room
temperature.